\documentclass[acus]{JAC2003}

\usepackage{graphicx}
\usepackage{booktabs}

\begin{document}
\title{CHAMBER SURFACE ROUGHNESS AND ELECTRON CLOUD FOR THE ADVANCED PHOTON SOURCE SUPERCONDUCTING UNDULATOR  \thanks{ Work supported by U.S. Department of Energy, Office of Science, Office of Basic Energy Sciences, Under Contract No. DE-AC02-06CH11357. Measurements taken on the Soft X-Ray Beamline at the Australian Synchrotron}}

\author{Laura Boon\thanks{ lboon@purdue.edu}, Purdue University, West Lafayette, IN, 47905, USA\\
Katherine Harkay, ANL, Argonne, IL 60439, USA}

\maketitle

\begin{abstract}
The electron cloud is a possible heat source in the superconducting undulator (SCU) designed for the Advanced Photon Source (APS), a 7-GeV electron synchrotron radiation source at Argonne National Laboratory.  In electron cloud generation extensive research has been done, and is continuing, to understand the secondary electron component.  However, little work has been done to understand the parameters of photoemission in the accelerator environment.  To better understand the primary electron generation in the APS; a beamline at the Australian Light Source synchrotron was used to characterize two samples of the Al APS vacuum chamber.  The total photoelectron yield and the photoemission spectra were measured.  Four parameters were varied: surface roughness, sample temperature, incident photon energy, and incident photon angle, with their results presented here.  
\end{abstract}

\section{Introduction}

The upgrade of APS calls for the production of a higher energy, higher-brightness photon beam~\cite{bib:CDR}.  To achieve this goal a new superconducting undulator (SCU) has been designed at APS.  This will be an out-of-vacuum undulator with a period of 1.8 cm.  The chamber in the SCU cryostat will be thermally isolated from the superconducting coils and kept at 20K, the chamber cooling will  be provided by two cryocoolers producing 40 W of cooling power.  

Studies performed on the in-vacuum superconducting undulator at ANKA, an electron synchrotron, confirm that their unexpected heat load is from electron cloud multipacting~\cite{bib:ANKA}.  However, work is still being done to find what mechanism is responsible for electron cloud multipacting, which is not reproduced in electron cloud simulation codes for electron machines.

\section{Simulations}

\begin{figure}[htb]
   \centering
   \includegraphics*[width=65mm]{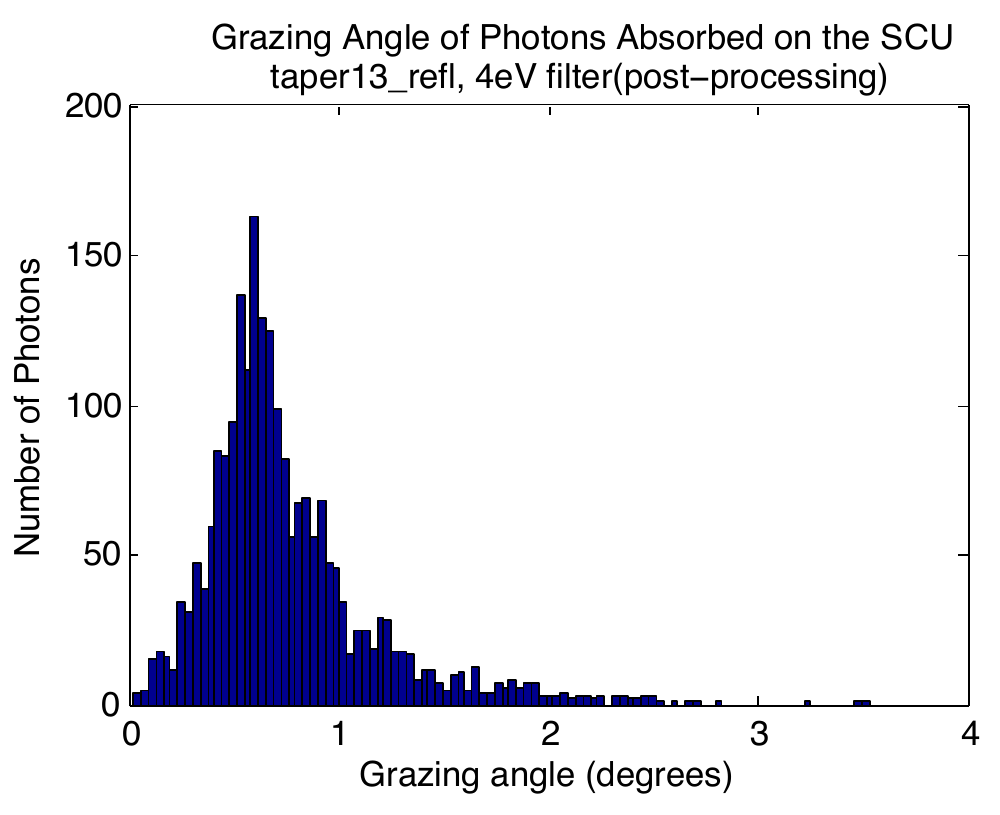}
   \caption{Incident grazing angle of absorbed photons in the SCU cryostat chamber.  Results produced from simulations using Synrad3d. }
   \label{fig:Abs_angle}
\end{figure}

\begin{figure}[htb]
   \centering
   \includegraphics*[width=65mm]{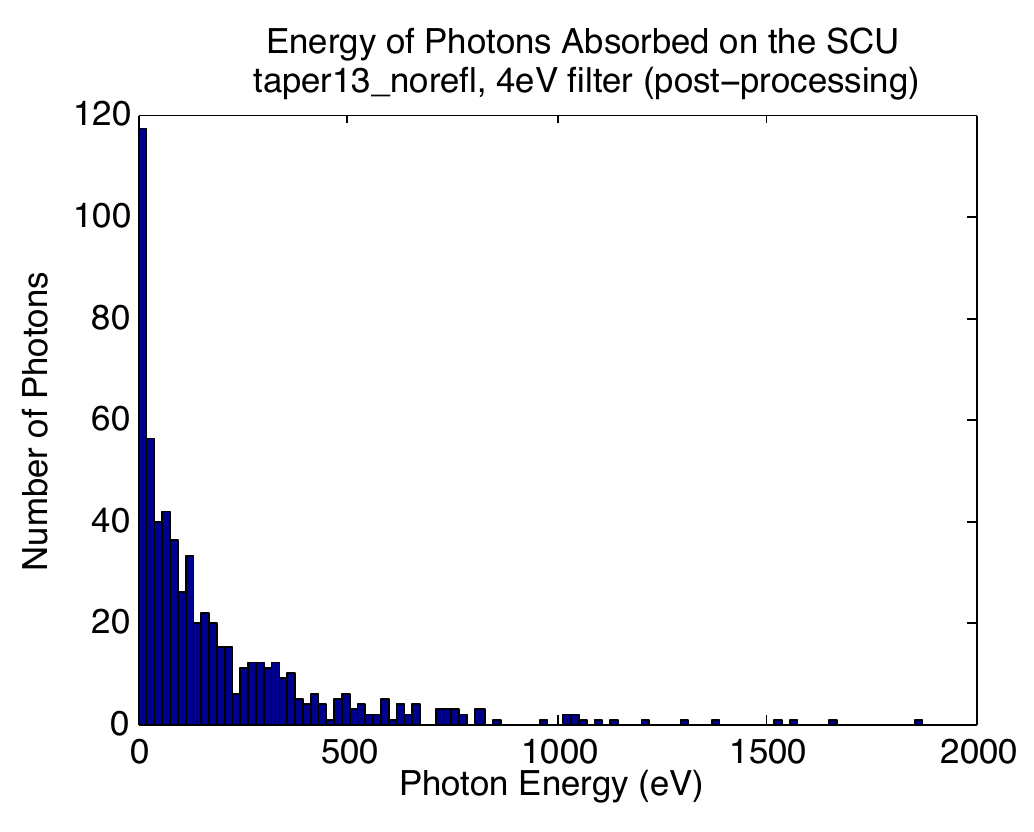}
   \caption{Energy of absorbed photons in the SCU cryostat chamber.  Results produced from simulations using Synrad3d. }
   \label{fig:Abs_energy}
\end{figure}

The range of photon angle and energy chosen for the measurement was based on simulations using synrad3d~\cite{bib:synrad3d}, in which a specular scattering model was assumed.  The peak flux of absorbed photons was at 0.6 degrees grazing angle, as shown in Figure~\ref{fig:Abs_angle}.  Although the critical energy from an APS bending magnet is 19 KeV the flux of photons absorbed on the chamber walls of the SCU is peaked at low energies, as shown in Figure~\ref{fig:Abs_energy}.  Due to physical limitations of the beamline the smallest grazing angle measurable was 3 degrees.

\section{Measurement Description}

The Australian Synchrotron's Soft X-ray beam line was used to measure the quantum efficiency of actual beam chamber samples.    Data were acquired at grazing angles of  3, 5, 10, and 50 degrees in photon energy scans from 100 eV to 2000 eV in 0.5 eV steps at varying temperatures from 300K to 180K.  

\begin{figure}[htb]
   \centering
   \includegraphics*[width=65mm]{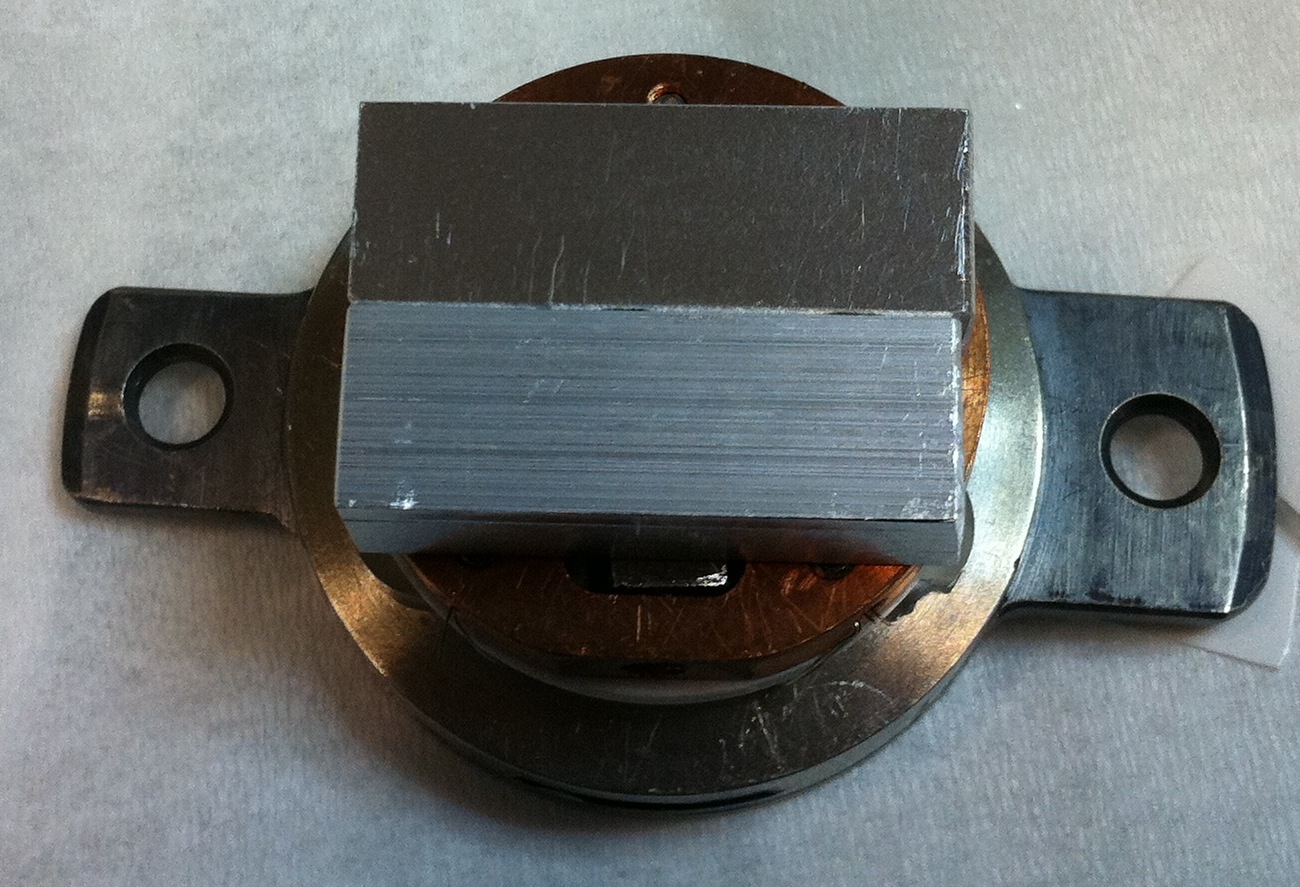}
   \caption{Picture of the sample holder with the Al samples.  The smooth, or polished sample is on the top, while the rough, or unpolished sample is on the bottom.}
   \label{fig:Sample_holder}
\end{figure}

Two samples were measured.  Both are sections of the extruded Al APS beam chamber, a rough or unpolished sample of the SCU Al chamber and a smooth or polished sample.  The smooth sample, the top sample in Figure~\ref{fig:Sample_holder}, was polished using an abrasive flow process~\cite{bib:trakhtenberg_2011}.  The rough sample, the bottom sample in Figure~\ref{fig:Sample_holder}, was measured "as received."  The surface roughness of the samples was measured and is presented in Table~\ref{table:Surface_roughness}.  The samples were cleaned in an ultrasonic, acetone bath for ten minutes before the measurements were taken.

\begin{table}[hbt]
   \centering
   \caption{Sample parameters}
   \begin{tabular}{lcc}
       \toprule
       \textbf{Sample} & \textbf{RMS}  \\ 
       \midrule
           Rough                            & 1180 nm               \\
           Smooth                           & 139 nm         \\
       \bottomrule
   \end{tabular}
   \label{table:Surface_roughness}
\end{table}

\section{Analysis}

At the beamline two different sample drain currents were measured.  To calibrate the total photon flux on the sample, the drain current from the Si diode at the back of the sample chamber was measured for all photon energies.  Using the data for the Si diode given by the manufacturer~\cite{bib:Sidiode_manufacturer}, the photon flux at each energy can be calculated. Equation~\ref{equ:photon_flux} was used to calculate the total photon flux incident on the Si diode.

\begin{equation}
Flux = \frac{I_{Si}}{q_e}\frac{3.65eV}{electron}\frac{1}{E_{photon}} F(E_\gamma) \left[\frac{photons}{sec} \right]
\label{equ:photon_flux}
\end{equation}

$I_{Si}$ is the drain current measured from the Si diode, 3.65 eV is the average energy for an electron-hole pair creation in silicon, $E_{photon}$ is the energy of the incident photon beam and $F(E_\gamma)$ is the transmission coefficient~\cite{bib:XrayCenter}.   With the sample in place all photons that would have been incident on the diode are now incident on the sample, therefore the flux on the sample is also given by equation~\ref{equ:photon_flux}.
The drain current from the sample was also measured. Using equation~\ref{equ:Electrons}, the number of electrons produced can be calculated.

\begin{equation}
\# Electrons = \frac{I_{Al}}{q_e} \left[ \frac{electrons}{sec} \right]
\label{equ:Electrons}
\end{equation}

$I_{Al}$ is the drain current measured from the aluminum sample, and $q_e$ is the charge of an electron.  The quantum efficiency is the ratio of number of electrons emitted to number of incident photons, as seen in equation~\ref{equ:QE}.

\begin{equation}
QE = \frac{ \# Electrons}{Flux}
\label{equ:QE}
\end{equation}

Equation ~\ref{equ:QE} was used to calculate the quantum efficiency for Figure ~\ref{fig:QE_5}.  All results were normalized to the storage ring beam current.

\section{Results}

\begin{figure}[htb]
   \centering
   \includegraphics*[width=65mm]{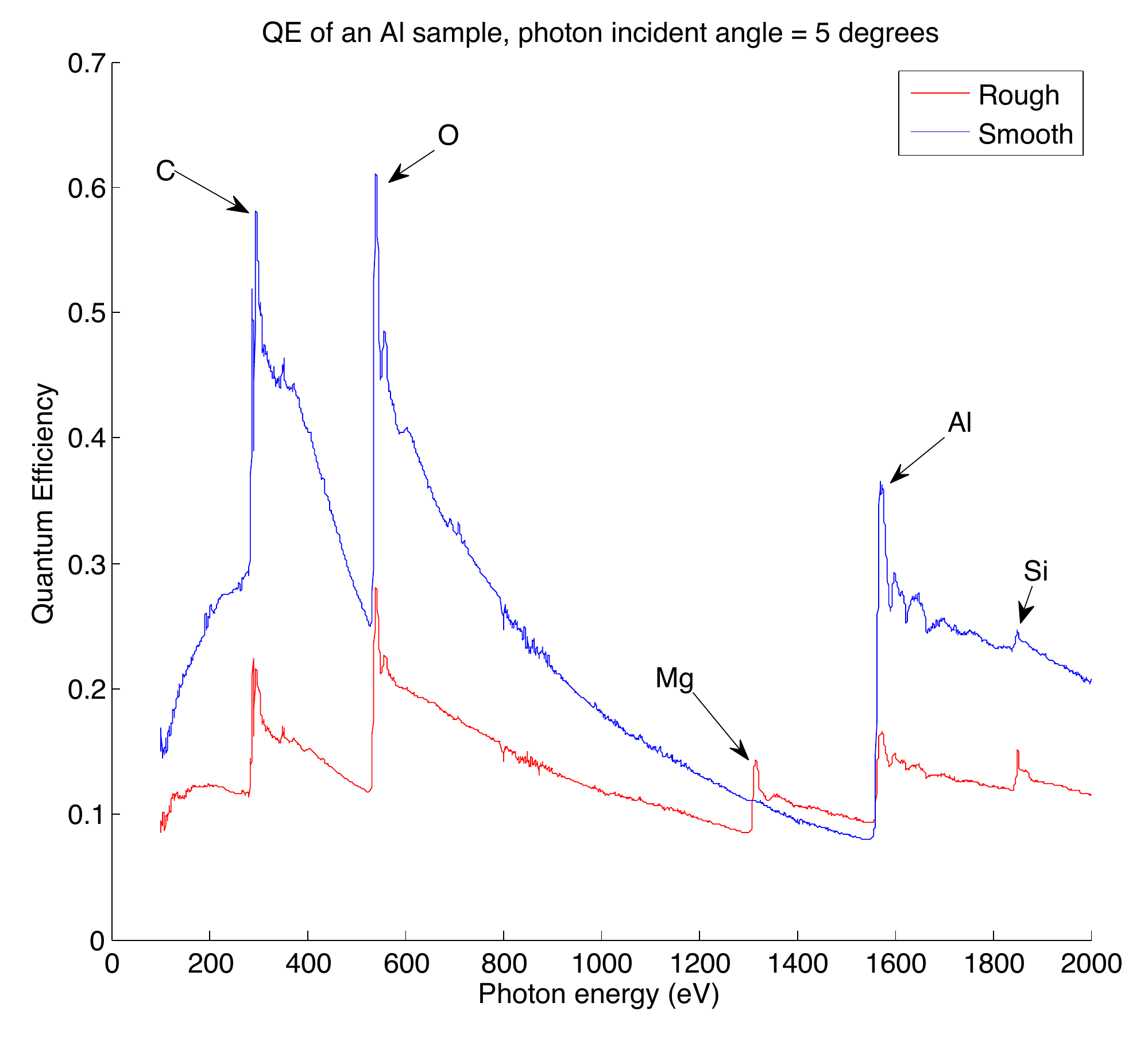}
   \caption{An example of a QE plot as a function of energy.  These data were taken with the incident photons at a grazing angle of 5 degrees. The sample was at 190 K.}
   \label{fig:QE_5}
\end{figure}

The quantum efficiency as a function of energy was found to be strongly dependent on the energy of the incident photon beam.  As an example, the quantum efficiency of the Al chamber with a photon beam at a grazing angle of 5 degrees is shown in Figure~\ref{fig:QE_5} for both samples.  There is an increase in the quantum efficiency for photon beam energies equal to the K and L edges of the materials.  Although the substrate does not contain any oxygen and carbon both are measured on the surface.  The oxygen and carbon are part of the aluminum oxide layer, typically 4-5 nm~\cite{bib:AlOx}, that forms on aluminum from exposure to the air.  The quantum efficiencies measured with low grazing angles 3-5 degrees indicated a stronger dependence on the surface chemistry, while the quantum efficiencies at larger angles, around 50 degrees, were more dependent on the substrate.

\begin{figure}[htb]
   \centering
   \includegraphics*[width=65mm]{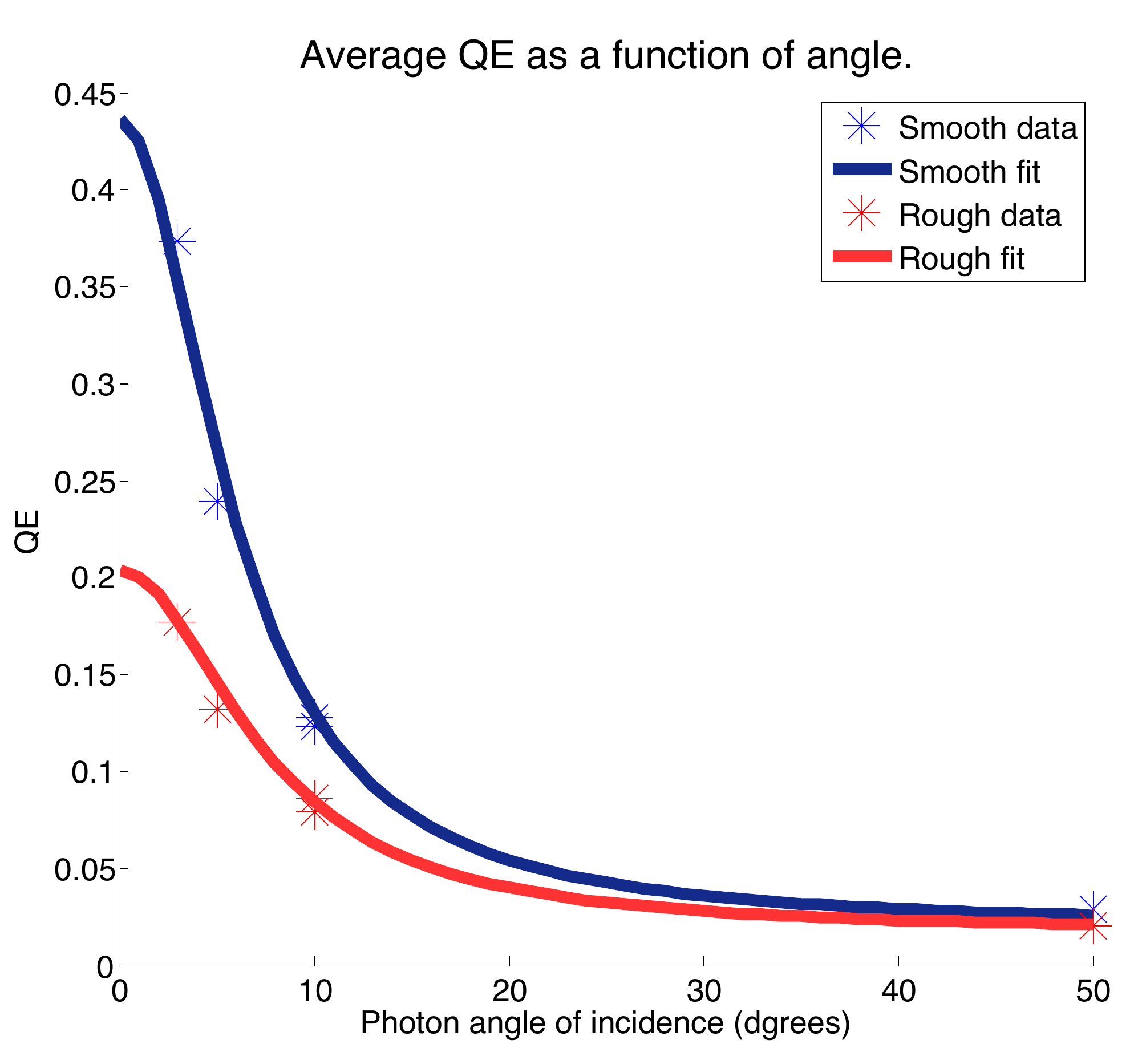}
   \caption{The average QE plotted as a function of photon grazing angle.  Higher photon grazing angles have a smaller QE than low grazing angles.}
   \label{fig:QE_average}
\end{figure}

\begin{figure}[htb]
   \centering
   \includegraphics*[width=65mm]{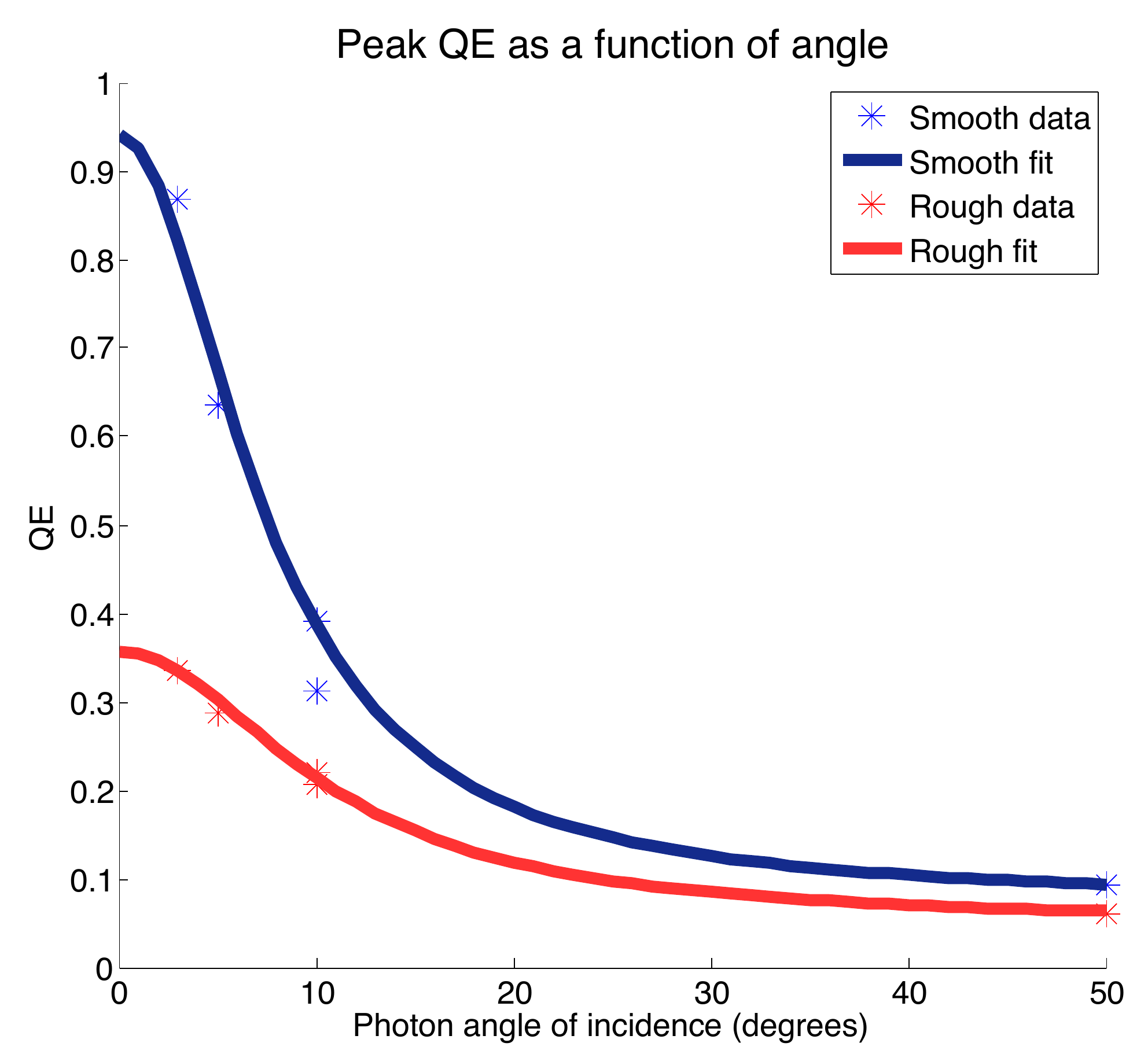}
   \caption{The peak QE plotted as a function of photon grazing angle.  Higher photon grazing angles have a smaller QE than low grazing angles.}
   \label{fig:QE_peak}
\end{figure}

The average quantum efficiency was calculated for each angle and sample and then fit to a Lorentzian, see Figure~\ref{fig:QE_average}.  Similarly the peak QE was found at each angle then fit to a Lorentzian, see Figure~\ref{fig:QE_peak}.

From Figures~\ref{fig:QE_average} and~\ref{fig:QE_peak} it is seen that the smoother sample had a higher quantum efficiency than the rough sample.  The difference is greater for smaller grazing angles  and reduces as the photon grazing angle approaches 50 degrees.  The angle dependance of the quantum efficiency is related to the penetration depth of photons.  For all grazing angles the photons travel the same distance through the material; however, the photons are absorbed closer to the surface when the grazing angle is low.  The escape probability of the photoelectrons is thereby increased, which increases the quantum efficiency.   

To use this data to update current electron cloud generation codes, the quantum efficiency for angles less than the measured three degrees would need to be interpolated from the data.  Figures~\ref{fig:QE_average} and ~\ref{fig:QE_peak} show a good fit with a Lorentzian for photon grazing angles between 3 and 50 degrees, and the quantum efficiency for less than three degrees would be interpolated from the fit.  The energy dependance of the quantum efficiency at photon grazing angles smaller than three degrees can also be determined.  Knowing that the surface chemistry will dominate, a higher quantum efficiency would be assumed for photons with energies equal to the K and L edges of carbon and oxygen.

\begin{figure}[htb]
   \centering
   \includegraphics*[width=65mm]{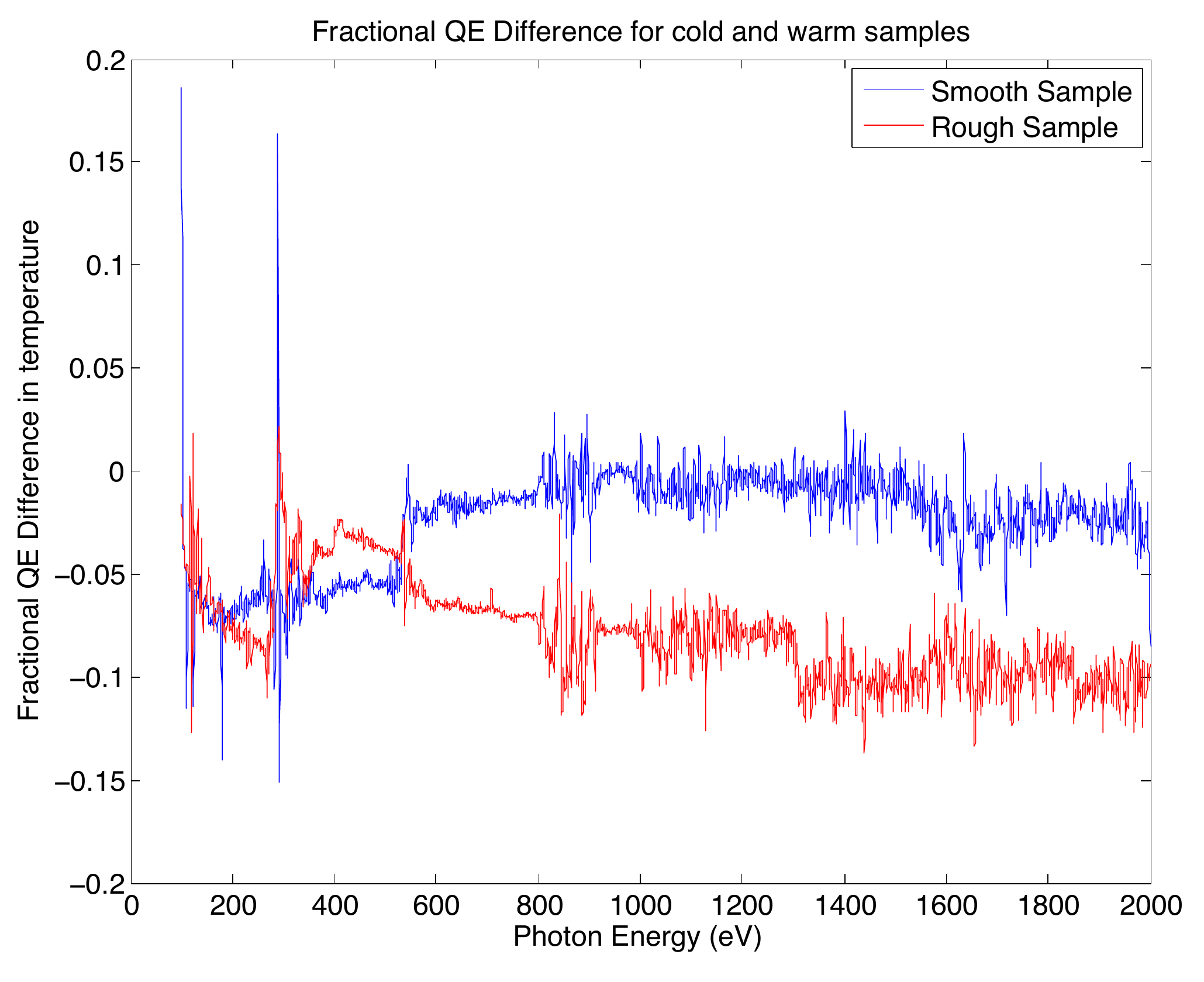}
   \caption{The difference in quantum efficiency for a cold (180K) and warm (300 K) sample.}
   \label{fig:QE_temp}
\end{figure}

For the results shown in Figures~\ref{fig:QE_average} and~\ref{fig:QE_peak} the temperatures of the samples was assumed to be constant.  This assumption is based on a comparison of the quantum efficiency at the same angle and different temperatures (300K and 180K).  The quantum efficiency over all energies did not vary more than 15\%.  For the superconducting undulator the chamber will be held at 20K, which will require more studies to determine if there will be a difference in quantum efficiency at the lower temperatures.  

\section{Conclusions}

The quantum efficiency was measured for the aluminum chamber that will be installed at APS for the SCU.  The quantum efficiency was seen to be strongly dependent on the surface roughness and grazing angle of the photon beam.  This study shows the actual quantum efficiency is more complex then a single value that is currently being used in electron cloud codes.  More work must still be done to understand the impact of these results on the SCU. 

\section{Acknowledgements}

Thanks to B. Cowie and K. Wootton for help in making the measurements.  To L. Assoufid, E. Trakhtenberg and G. Wiemerslage for preparing the samples.  To R. Cimino, S. Casalbuoni, L. Emery and B. Yang for useful discussions and analysis help and M. Harrison for ILC financial support.

\end{document}